\documentclass[aps,prd,english,superscriptaddress,11pt,notitlepage]{revtex4}
	\usepackage[colorlinks=true, a4paper=true, pdfstartview=FitV,
linkcolor=blue, citecolor=blue, urlcolor=blue]{hyperref}

\usepackage{xcolor}
\usepackage{graphicx}
\usepackage{dcolumn}
\usepackage{bm}
\usepackage[colorlinks=true, a4paper=true, pdfstartview=FitV,
linkcolor=blue, citecolor=blue, urlcolor=blue]{hyperref}

\begin{document}

\vspace*{-1.0cm}

\title{Sphalerons and resonance phenomenon in kink-antikink collisions}

\author{C. Adam}
\affiliation{
Departamento de Fisica de Particulas, Universidad de Santiago de Compostela
and Instituto Galego de Fisica de Altas Enerxias (IGFAE), E-15782 Santiago de Compostela, Spain
}

\author{D. Ciurla}

\author{K. Ole\'{s}}%

\author{T. Roma\'{n}czukiewicz}

\author{A. Wereszczy\'{n}ski}

\affiliation{
 Institute of Theoretical Physics,  Jagiellonian University, Lojasiewicza 11, 30-348 Krak\'{o}w, Poland
}

\date{\today}

\begin{abstract}
We show that in some kink-antikink (KAK) collisions sphalerons, i.e., unstable static solutions - rather than the asymptotic free soliton states - can be the source of the internal degrees of freedom (normal modes) which trigger the resonance phenomenon responsible for the fractal structure in the final state formation. 
\end{abstract}

\pacs{Valid PACS appear here}
\maketitle

\section{\label{sec:intro}Introduction}
Nonlinear field theories supporting kinks in one plus one dimension, and their embeddings in higher dimensions called defects, have a wide range of applications, from protein folding and optical fibers to particle physics and cosmology. Surprisingly, the scattering of kinks reveals a fascinating and rather complex  behavior \cite{SM, Shnir, Kev}, despite its conceptual simplicity.  The best example is the fractal structure observed in the final state formation in KAK scattering in non-integrable scalar field theories like the $\phi^4$ model \cite{Sug, Mosh, CSW, Goodman, TW} where, depending on the initial velocity $v_{i}$ of the colliding solitons, two main channels occur: {\it 1)} annihilation to the vacuum via the formation of quasi-periodic, slowly decaying states ("bions"), or {\it 2)} the reappearance of free kinks after a few {\it bounces}. Both the regions where a bion is created (bion chimneys) and the $n$-bounce regions (bounce windows) show a fractal pattern.

Qualitatively, the existence of the fractal structure is attributed to the so-called {\it resonance phenomenon} \cite{Sug}, \cite{CSW},  \cite{TW} which couples the kinetic degrees of freedom (DoF) of a soliton with the internal ones. The kinetic (translational) DoF are obviously related to the lightest excitation of the soliton, its zero mode. The internal DoF are typically the massive normal (or quasi-normal \cite{trom}) modes of the asymptotic states, i.e., of the free, infinitely separated solitons. During a collision, the energy which is initially stored in the kinetic DoF of the incoming solitons (and possibly also in initially exited massive modes) may be transferred temporarily to the internal DoF, which results in a complicated pattern of final states. 

Here it is assumed that the normal modes take the fixed form resulting from the asymptotic kink and antikink during the whole scattering, and no deformation due to the presence of the collision partner is taken into account. We say that the normal modes are "frozen". The reason is that for KAK configurations with a finite separation no static solution with a well-defined linear problem (i.e., well-defined normal modes) is available in the $\phi^4$ model. For many years it was believed that the non-integrable mechanical system resulting from a  "collective coordinate" (CC) approximation, where only the relative KAK separation $s(t)$ and the two normal-mode amplitudes $A_1(t)$ and $A_2(t)$ are promoted to dynamical variables, allows for a reliable quantitative description of KAK scattering and the resonance mechanism. A typographical error \cite{TW} in the original paper \cite{Sug} which propagated through the subsequent literature, however, invalidated those conclusions. Only very recently it has been shown that the CC approach allows for a reliable description of KAK scattering in the reflection-symmetric case $A\equiv A_1 = - A_2$  \cite{MORW}. Here the two crucial steps are {\em i)} the removal of a coordinate singularity in the "moduli space" spanned by the CC, and {\em ii)} a judicious choice of the initial value of the amplitude $A$, which must be inferred from the one-kink sector.  The same resonance mechanism is assumed to be responsible for the existence of fractal structures also in many other models \cite{kiv, GH, izquierdo, long-tail, simas, bazeia, yan, Moh, villa, Sangmo}.  

In contrast, in some theories there exist static KAK configurations solving the static Euler-Lagrange equations for arbitrary KAK separation, implying the existence of (at least) a one-parameter family of static KAK solutions, parametrized, e.g., by the separation parameter $s$. In this case, each static KAK solution has its own linear problem, and the resulting normal modes and their frequencies, in general, vary with $s$. Adiabatic low-energy KAK scattering may be described by an effective motion on the subspace spanned by $s$, but for more excited initial conditions the normal modes and their variation with $s$ play a crucial role. In particular, whenever a discrete normal mode frequency disappears into the continuum at a particular value $s_{\rm SW}$ of $s$, this has a huge impact on the scattering process (spectral wall phenomenon \cite{spectral-wall, no-force-scatt, two-field-SW}). In an effective CC description, it is mandatory to include the full "dynamical" normal modes, i.e., their variation with $s$. Any restriction to the frozen normal modes completely misses the spectral walls.

In this letter, we consider an intermediate situation, where in addition to the free infinitely separated solitons there exists a {\it sphaleron} \cite{T, KM, MSam}, i.e., an unstable static KAK solution at a certain finite separation $s_b$. It turns out that the lowest normal mode of this unstable solution can be the main factor governing the dynamics of KAK collisions in some cases.  Specifically, this normal mode may act as the internal DoF capable of storing energy and triggering the fractal structure of final states.
This result presents another manifestation of the importance of the temporary form of the field and, therefore, of the temporary properties of the related normal modes, in KAK collisions.  More generally, our results demonstrate the high significance of sphalerons in the dynamics of topological solitons.

\section{\label{sec:model}Model with unstable solution}

The family of models which we consider represent a modification of $\phi^4$ theory 
\begin{equation} \label{L-nsd}
L[\phi, \sigma; \epsilon]=(1-\epsilon^2) L[\phi,\sigma]+ \epsilon^2 L [\phi],  
\end{equation}
where
\begin{equation}
 L[\phi]=\frac{1}{2} \int_{-\infty}^\infty  dx \left[ \phi_t^2- 
\phi_x^2  - (1 - \phi^2)^2    \right] \label{phi4}
 \end{equation}
 is the usual $\phi^4$ model, while  
\begin{equation}
L[\phi, \sigma]=\int_{-\infty}^\infty  dx \left(  \frac{1}{2}\phi_t^2  - \frac{1}{2}
\left( \phi_x  + \sigma (1-\phi^2 ) \right)^2 \right)
\end{equation}
is a BPS-impurity modification of the $\phi^4$ model, where the impurity (background field) $\sigma = \tanh (x)$ \cite{solvable-imp, no-force-scatt}, and $\epsilon \in [0,1]$. 
Here BPS (=Bogomolnyi-Prasad-Sommerfield) refers to the fact that the static field equations can be reduced to first-order ("BPS") equations, which possess a one-parameter family of KAK solutions, see eq. (\ref{BPS-sol}) below.

We choose the particular model (\ref{L-nsd}) because it has an unstable static solution - a sphaleron -  for any $\epsilon \in (0,1)$. Moreover, it allows to study the competition between the standard mechanism, where the frozen shape mode of the free soliton takes part in the resonant phenomenon, and the new mechanism due to the unstable static solution. The former occurs as $\epsilon \to 1$, as in the limit $\epsilon =1$ we recover $\phi^4$ theory, while the latter  dominates for $\epsilon^2 \lesssim 0.4$.

\begin{figure}
\hspace*{-1.0cm}
\includegraphics[width=8.5cm]{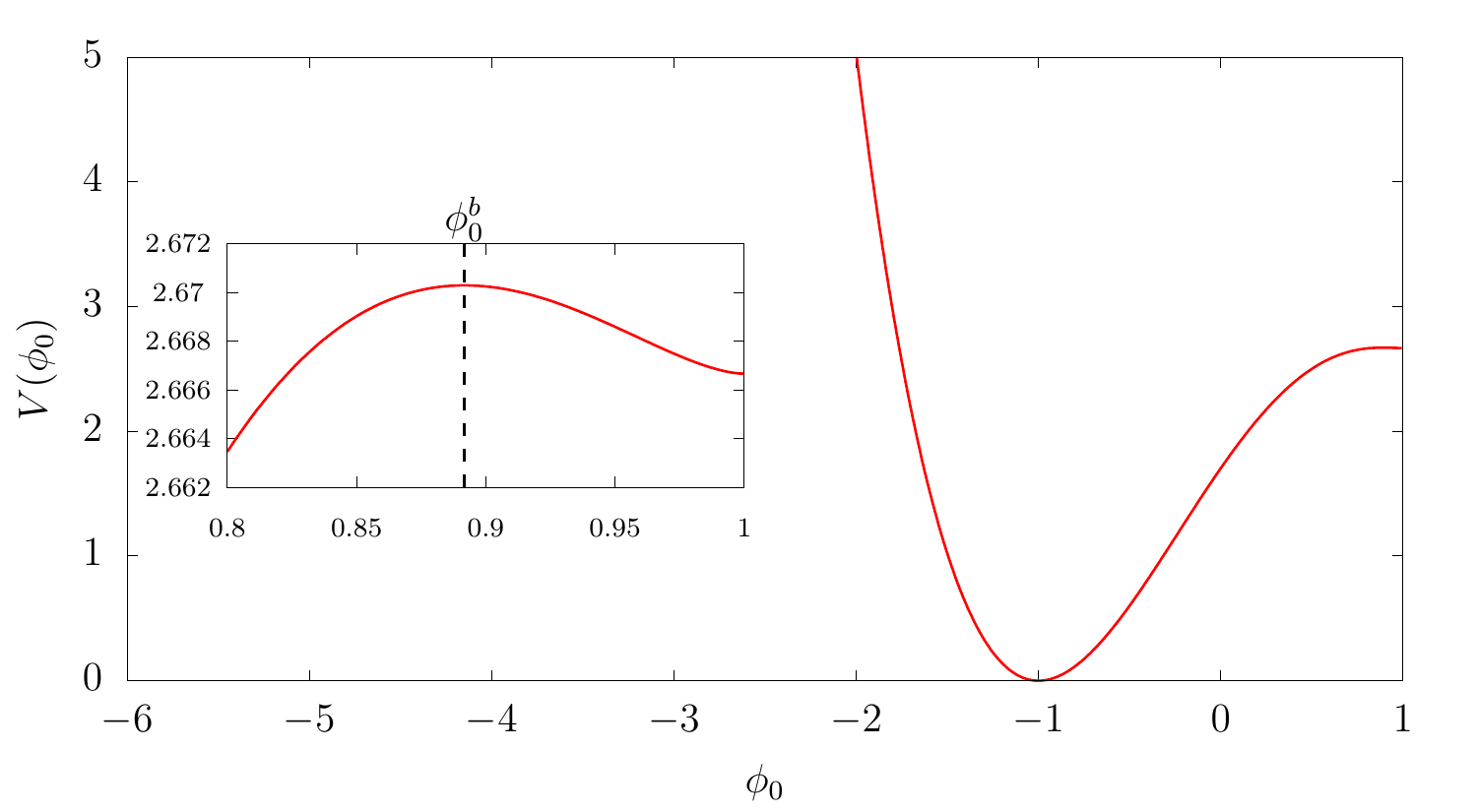}
\includegraphics[width=8.7cm]{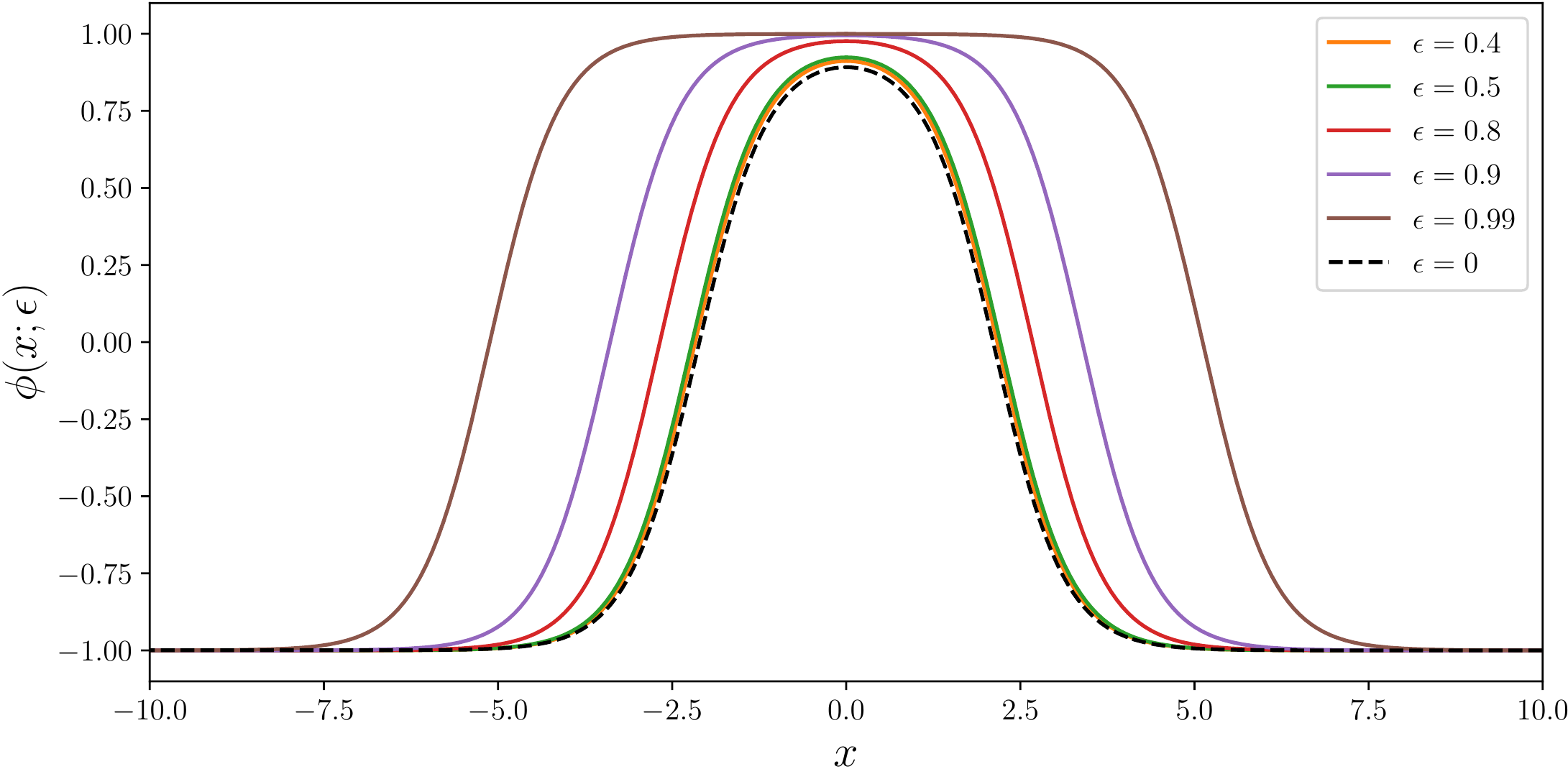}
\caption{Left: the effective potential $ V(\phi_0)$. Right: the sphaleron for various $\epsilon$}
\label{fig-potential}
\label{unstable}
\end{figure}

To understand the structure of solutions, let us begin with the $\epsilon=0$ limit, where there are infinitely many static, energetically equivalent solutions \cite{no-force-scatt}
\begin{equation}
\phi(x; \phi_0) = \frac{(1+\phi_0) -(1-\phi_0)\cosh^2 x}{(1+\phi_0) +(1-\phi_0)\cosh^2 x}. \label{BPS-sol}
\end{equation}
Their energy is set to 0. They are parametrized by $\phi_0 \in (-\infty,1)$, which coincides with the value of the field at the origin, and describes a kink and antikink at arbitrary distance \cite{footnote1}. For example, for $\phi_0 \to 1$ we arrive at the infinitely separated KAK pair while for $\phi_0=-1$ the solitons lie on top of each other, giving the $\phi=-1$ vacuum. 
This degeneracy of the KAK solutions (\ref{BPS-sol}) is lifted if $\epsilon \neq 0$ by the following static (potential) energy emerging from (\ref{phi4})
\begin{equation}
V(\phi_0)=\epsilon^2  \int dx \frac{1}{2} \left( [\phi_x (x;\phi_0)]^2 + [1-\phi^2 (x;\phi_0)]^2 \right). \label{eff-pot-eps}
\end{equation}
Now, the ground state is $\phi=-1$ with $V=0$, while the infinitely separated KAK pair is a local minimum, where $V=8/3$. Importantly, there is also a local maximum, occurring at $\phi_0^b\approx 0.89167$, which suggests  the existence of an unstable solution, a sphaleron, see Fig. \ref{fig-potential}. Indeed, we find this solution numerically for any $\epsilon$ by solving the static Euler-Lagrange (EL) equation resulting from (\ref{L-nsd}), see Fig. \ref{unstable}. For $\epsilon^2  \lesssim 0.4$ they are very well approximated by the exact solution $\phi(x; \phi_0^b)$ of eq. (\ref{BPS-sol}). 

Concerning the spectrum of small perturbations, we first observe that the continuum threshold is $\epsilon$ independent, $\omega^2_{c}=4$ \cite{small-force-scatt}. 
As indicated, there are two configurations, i.e. (asymptotic) static solutions, with a well-defined linear problem, namely the asymptotic KAK pair and the sphaleron.  
The asymptotic state is always an infinitely separated $\phi^4$ KAK pair, separated by the vacuum $+1$,
\begin{equation}
\phi(x) = \tanh(x+x_0)-\tanh(x-x_0)-1, \label{KAK}
\end{equation}
where $x_0\to \infty$. Therefore, each soliton has one massive normal mode, i.e., the {\it shape mode}, $\eta(x)=\sinh(x\pm x_0)/\cosh^2(x\pm x_0)$ with frequency $\omega^2_{shape}=3$. The vacuum $+1$ also provides two normal modes located at the origin, as a consequence of the presence of the impurity $\sigma$. Their frequencies grow with $\epsilon$. Specifically, $\omega^{2}_{imp \; (n)} = 4-1/4(\sqrt{25-24\epsilon^2}-1-2n)^2$, $n=0,1$. They disappear into the continuum spectrum at $\epsilon^2=1$ and $\epsilon^2=2/3$ for $n=0,1$ respectively, as it must be (they must be absent in the no-impurity limit $\epsilon \to 1$). The sphaleron solution, on the other hand, contributes with an obvious unstable mode with $\omega_0^2 <0$ and several positive frequency normal modes. In Tab. I we list the frequency $\omega_2$ of the most important, first even normal mode for several $\epsilon$.

\section{\label{sec:origin} Flow of the fractal structure}
\begin{figure}
\hspace*{-0.5cm} \includegraphics[width=1.05\columnwidth]{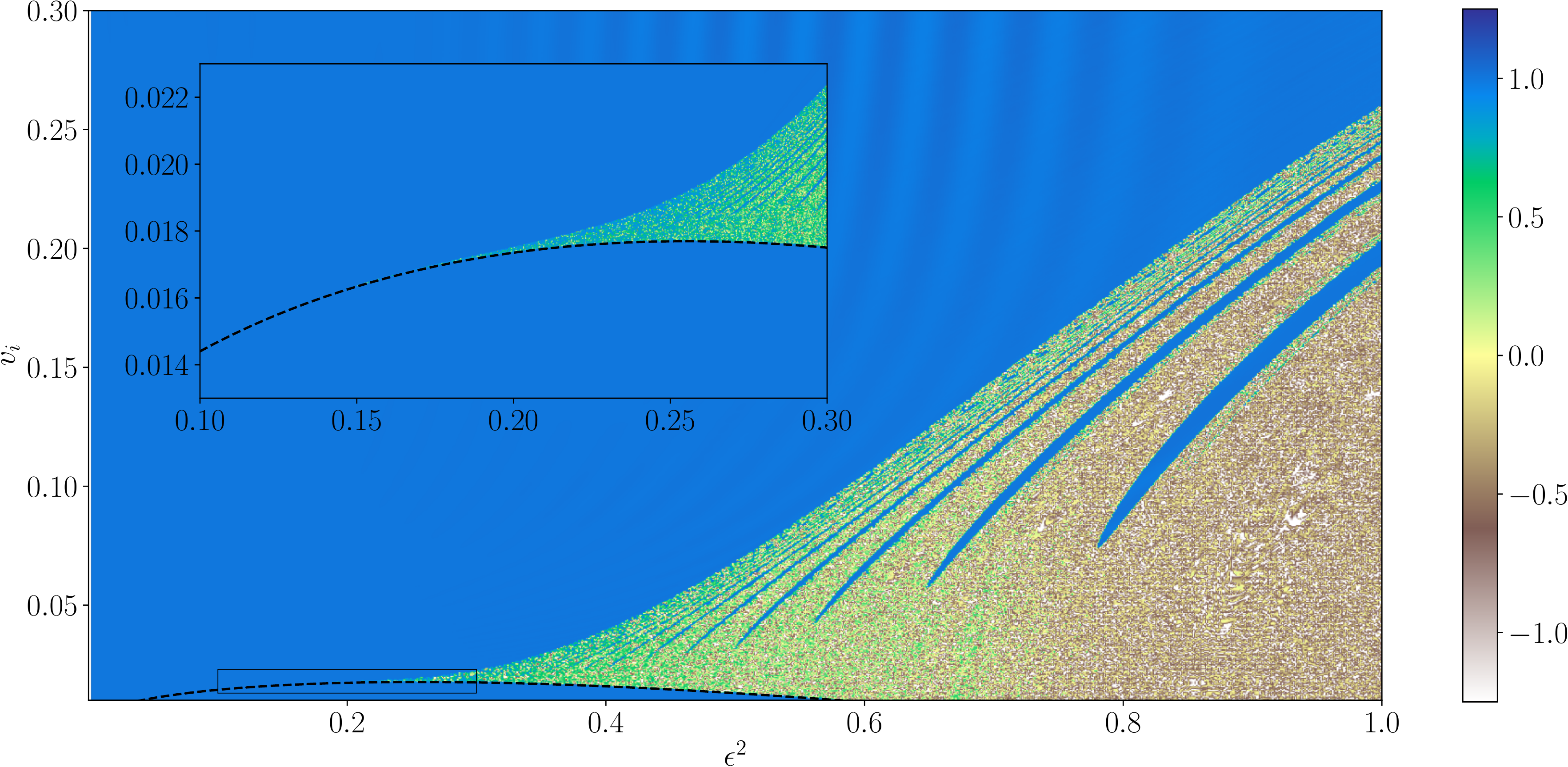}
 \caption{Flow of the fractal structure with $\epsilon$. The dashed line represents $\tilde{v}_{crit}$.}\label{fig:fractal}
\end{figure}

Now we collide the solitons. We choose a largely separated KAK pair as our initial configuration, which are boosted towards each other
\begin{equation}
\phi_{in}(x,t) =  \tanh(\gamma(x-v_{i}t+x_0))-\tanh(\gamma(x+v_{i}t-x_0)) -1. \label{init}
\end{equation}
Here, $2x_0$ is the separation distance, $v_{i}$ is the initial velocity of the solitons and $\gamma=(1-v_{i}^2)^{-1/2}$. This is numerically evolved in the EL equation for $\epsilon \in [0,1]$. For each model (each $\epsilon$) we vary the initial velocity $v_i$ from 0 to 0.3. In Fig. \ref{fig:fractal} we plot the value of the field at the origin, $\phi(0,t)$, after a large time $t^*=550+40/v_{i}$ as a function of both the initial velocity $v_{i}$ and $\epsilon$. This shows how the fractal structure in the final state formation changes with $\epsilon$. 

In Fig. \ref{fig:fractal} the blue regions correspond to the case where the kink and antikink reappear in the final state and therefore represent 
bounce windows. For example, the big blue region is one-bounce scattering. There is also a zero-bounce window - a blue region (for very small $v_{in}$, below the multicolor region) related to an 
elastic process resulting from the existence of the unstable static solution and the corresponding local energy 
maximum which, during the evolution, acts as a kind of barrier. Initially separated solitons must have enough kinetic energy 
to go through this point. Therefore, there exists a critical velocity $\tilde{v}_{crit}$ below which the kinks cannot pass over the sphaleron. The line $\tilde{v}_{crit}$ can be found from the energy conservation law. As $\epsilon \to 1$, the sphaleron looks like a kink-antikink pair with the 
intersoliton distance going to infinity. Thus, its impact on the collision becomes immaterial, leading to a vanishing $\tilde{v}_{crit}$ as we flow to the $\phi^4$ model. 

\begin{figure}
\hspace*{-1.0cm} \includegraphics[width=19.1cm]{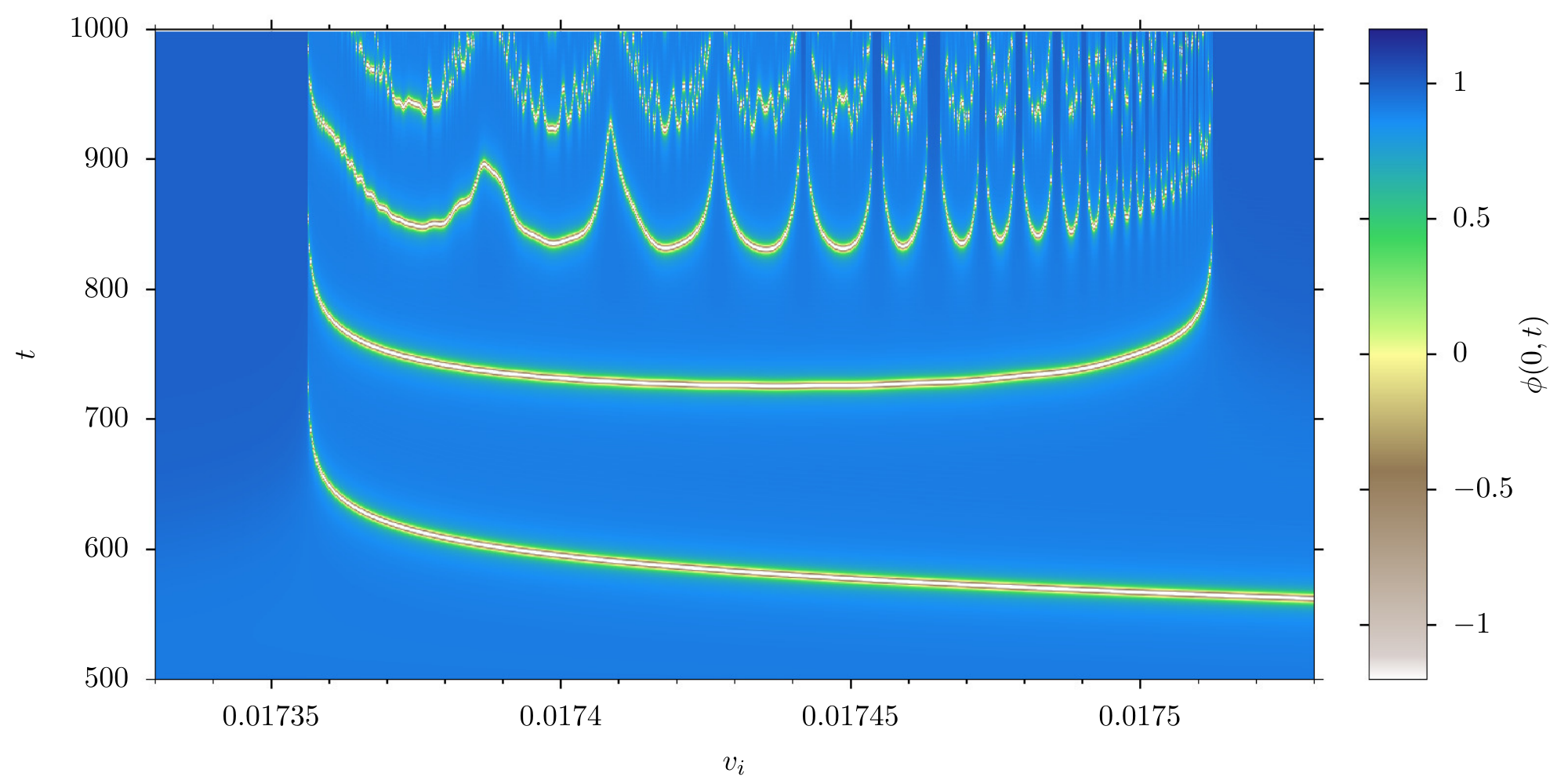}
\hspace*{-1.0cm} \includegraphics[width=19.1cm]{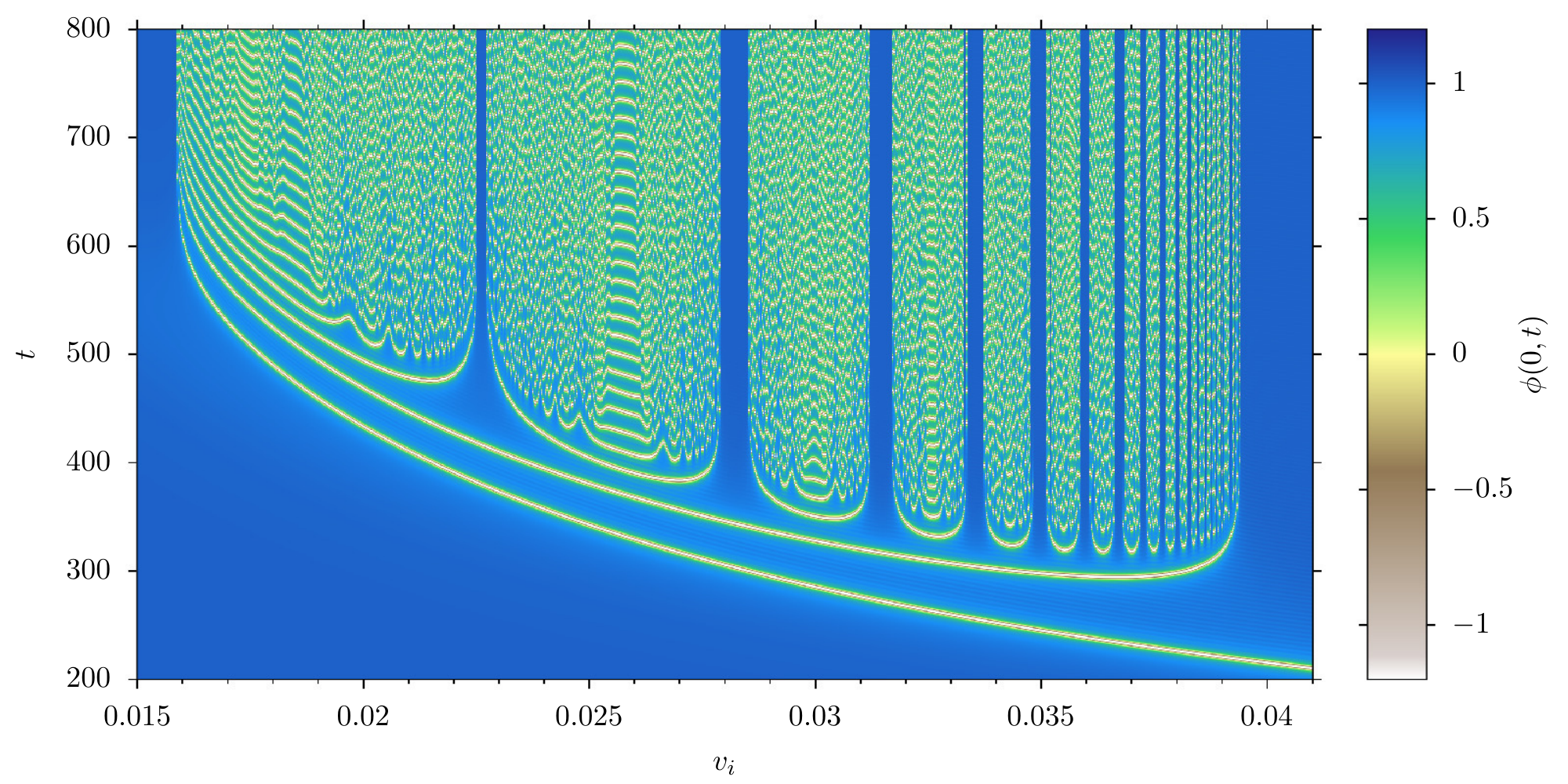}
\caption{Structure of the final states for $\epsilon^2=0.2$ (upper) and $0.4$ (lower)} 
\label{flow}
\end{figure}

Next, the multicolor (green, yellow and brown) regions correspond to bion chimneys in which a bion, a sort of quasi-oscillating state, decays into the vacuum via the emission of radiation. Typically, one defines another 
critical initial velocity, $v_{crit}$, at which the fractal structure accumulates, while above $v_{crit}$ only one-bounce scattering is 
observed. In Fig. \ref{fig:fractal} this velocity is defined by the border between the large blue and multicolor regions.

The two-bounce windows, bion chimneys and critical velocities are clearly visible in Fig. \ref{flow}, where we display the time dependence of the value of the field at the origin, $\phi(0,t)$, as a function of the initial velocity $v_{i}$ for $\epsilon^2=0.2$ and $0.4$. They are qualitatively similar to $\phi^4$ theory ($\epsilon=1$) although the range of $v_{i}$ at which the fractal structure exists strongly decreases as $\epsilon$ takes smaller values. Specifically, $v_{crit}$ decreases if $\epsilon \to 0$. One may observe that there are in fact two regimes: {\it (i)}  a linear decrease of $v_{crit}$ for $ 0.5 \lesssim \epsilon^2 \leq 1$; {\it (ii)} a much slower decrease of $v_{crit}$ for $ 0 \leq \epsilon^2 \lesssim 0.4$.  These two regimes reflect two different origins of the resonance phenomenon. 

\vspace*{-0.2cm}
\section{\label{sec:fractal} Origin of the resonance phenomenon}
\begin{figure}
\includegraphics[width=0.9\textwidth]{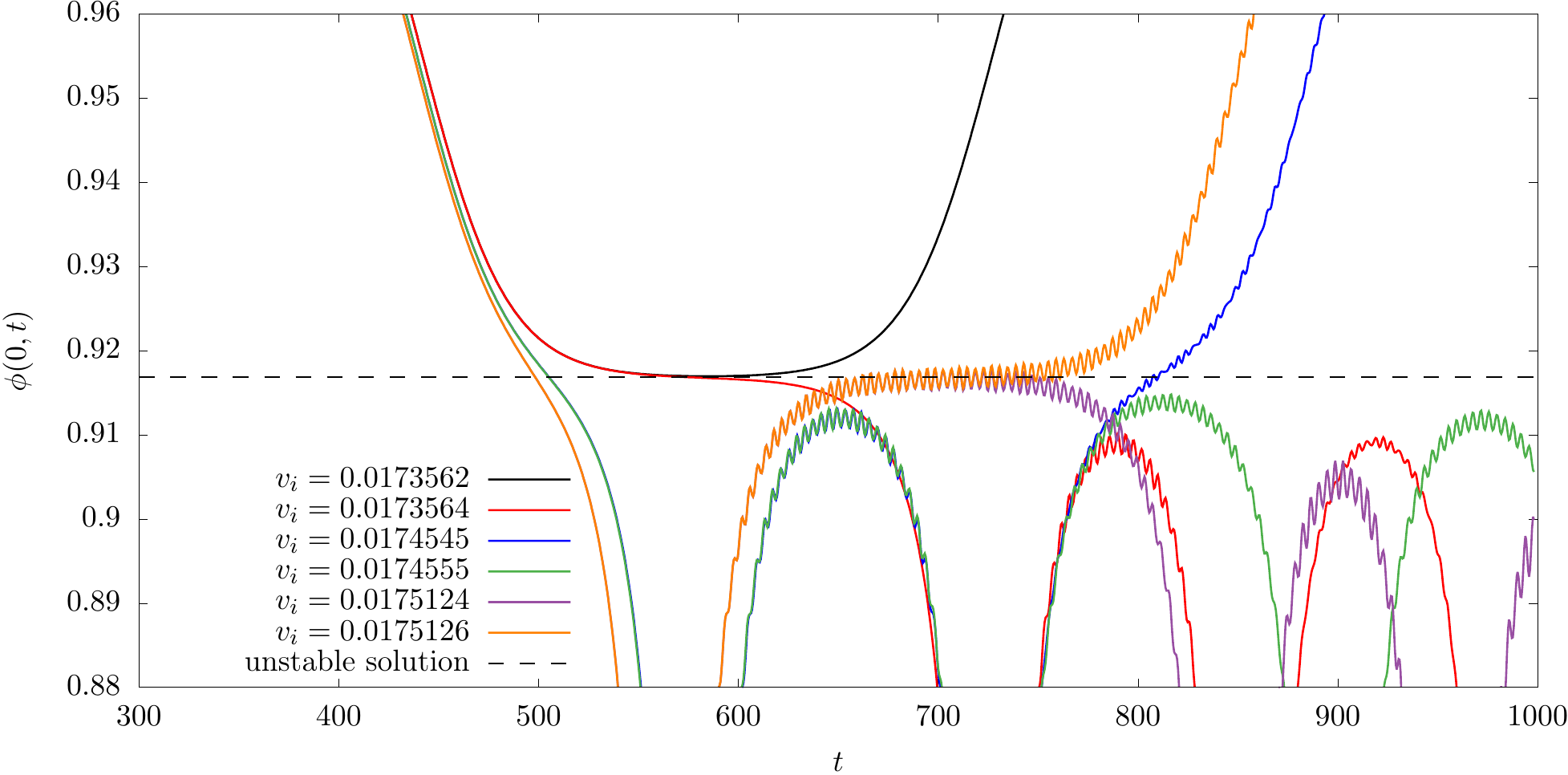} 
\caption{Bounces from the unstable solution for $\epsilon^2=0.2$.}\label{trajectories}
\end{figure}
Let us now concentrate on the second regime where $\epsilon^2 \lesssim 0.4$ and the sphaleron is very well approximated  by the exact configuration $\phi(x; \phi_0^b)$. 

To understand the bouncing windows in Fig. \ref{flow}, we take a closer look at the time dependence of the field at the origin $\phi (x=0, t)$. In Fig. \ref{trajectories} we show representative trajectories $\phi (x=0, t)$ for the $\epsilon^2=0.2$ case. The black curve shows KAK scattering with initial velocity $v=0.0173562$, that is, below the critical velocity $\tilde{v}_{crit}$. The result is the elastic (no bounce) reflection at the sphaleron (denoted as the dashed line). As the velocity increases, we enter the region of annihilation via bion formation. Indeed, for $v=0.0173564$ (red curve), the trajectory passes the unstable solution. Then, it is reflected at the repulsive core (repulsive interaction due to the potential (\ref{eff-pot-eps})), which excites a bound mode. However, the reflected solution has too little energy to climb once again through the sphaleron. As a result, the KAK solution oscillates between the core and the unstable maximum, slowly decaying into the $\phi=-1$ vacuum. For $v=0.0174545$, we explore one of the 2-bounce windows. The corresponding trajectory (blue curve) bounces twice at the repulsive core and once at the unstable maximum. Each bouncing at the core affects the amplitude of the bound mode, which finally allows the trajectory to pass the unstable barrier. Thus, in the final state we have a kink-antikink pair. For $v=0.017555, v=0.0174524$ (green and purple curves, respectively) we have again an annihilation regime via a sequence of reflections between the repulsive core and the unstable maximum. Note that for the purple trajectory, the solution stays at the unstable maximum for a relatively long time, forming a stationary oscillating state. In fact, this curve represents a situation very close to the border between annihilation and 2-bouncing windows. For example, as we very slightly increase the velocity to $v=0.0175126$ (orange curve), the stationary solution chooses to pass the maximum, and the 2-bounce solution emerges. We have seen that in the bounce windows the trajectories spend a long time oscillating on the sphaleron until they fall on one side or another. This choice is synchronized with the phase of the oscillations.
The same pattern repeats for bigger $\epsilon^2 \lesssim 0.4$. To conclude, the fractal structure in this regime is triggered by the sphaleron.

This is further confirmed by the observation that the oscillation frequency is the frequency of {\it a bound mode of the sphaleron} which significantly differs from the frequency of the shape mode, $\omega_{shape}=\sqrt{3}$, of the free solitons. Indeed, the standard fit of the dependence of the window duration $T$ and the window number $n$ 
\begin{equation}
 T = \frac{2\pi}{\omega_{res}}n+\delta
\end{equation} 
leads to the resonant frequency $\omega_{res}$ which, for sufficiently small $\epsilon$, is exactly the frequency of the first normal mode of the sphaleron, see Tab I. To conclude, in this regime of $\epsilon$, the duration of a 2-bounce solution in the n-th window is related to a bound mode excited on the unstable solution, and has nothing to do with a bound mode of the free (anti)kink.

\begin{table}  \label{tab-f}
{\scriptsize 
\begin{tabular}{ccc} 
\hline\hline
\hspace*{1cm}$\epsilon^2$\hspace*{1cm} &\hspace*{1cm} $\omega_{res}$ \hspace*{1cm} & \hspace*{1cm} $\omega_{2}$ \hspace*{1cm}  \\
\hline
\texttt{0.2} & \texttt{1.553} & \texttt{1.558} \\
\texttt{0.3} & \texttt{1.572} & \texttt{1.582}  \\
\texttt{0.4} & \texttt{1.576} & \texttt{1.609} \\
\hline\hline
\end{tabular}} 
\caption{Comparison of the fitted resonant frequency $\omega_{res}$ with the first even normal mode frequency $\omega_{2}$ of the sphaleron.}
\end{table}

In the regime where $\epsilon \to 1$, the well-know situation occurs. Namely, the free solitons provide the internal modes which enter into the resonance phenomenon, as very recently proved in \cite{MORW}. In the intermediate regime, both sources of the internal modes contribute, rendering the analysis quite involved. This regime requires further studies. 
\vspace*{-0.2cm}
\section{\label{sec:summary}Summary}
In this work, we have shown that the internal mode participating in the resonance phenomenon, which stands behind the fractal structures in the formation of the final state in KAK collisions, may have a completely different origin than usually assumed. In particular, such a mode can be provided by an unstable static solution, a sphaleron, rather than the stable, free single (anti)kink solution. We want to emphasize that the important property of the specific example we chose is the existence of the sphaleron  and {\em not} the presence of the impurity. The impurity model just provides a particularly simple realisation of this mechanism.  

Our results also indicate that the full understanding of soliton dynamics cannot be achieved only in terms of the asymptotic states, i.e., free solitons and their excitations \cite{nick}, even though such a frozen approach works well in many cases, see, e.g.,  \cite{MORW}. On the contrary, the temporary properties of the field can play the most prominent role and can, in fact, govern the KAK scattering. Obviously, this leads to an increased complexity of soliton dynamics and should result in the discovery of fractal structures in previously unexpected situations. The importance of transient, temporary states and their normal modes for KAK scattering was already emphasized in \cite{dor1} for the $\phi^6$ model, although no unstable solution was available in that case. 

Moreover, our results explicitly show the high relevance of sphaleron solutions in the nonlinear dynamics of kinks. An important role of sphalerons in the dynamics of nonlinear field theories was conjectured since their discovery. An explicit demonstration, however,  is in general difficult because of the complex nature of time-dependent problems in those theories.  
We would like to remark that unstable kinks (or domain walls) are common in models with two (and more) scalar fields as e.g., the Montonen-Sarker-Trullinger-Bishop model \cite{M, STB, Izq-2}. They also frequently exist in kink models with background fields where, at some point, the attractive force of a KAK pair can be compensated by external fields \cite{unst-osc} or by the presence of boundaries \cite{unst-bond-1}, \cite{unst-bond-2}. Finally, they have been recently found in a model with a higher order derivative term \cite{kev}.

\section*{Acknowledgements}
C.A and A.W. acknowledge financial support from the Ministry of Education, 
Culture, and Sports, Spain (Grant No. FPA2017-83814-P), the Xunta de 
Galicia (Grant No. INCITE09.296.035PR and Conselleria de Educacion) and 
the Spanish Consolider Program Ingenio 2010 CPAN (Grant No. 
CSD2007-00042). 
Further, this work has received financial support from Xunta de Galicia 
(Centro singular de investigación de Galicia accreditation 2019-2022), by 
the European Union ERDF, and by the “María de Maeztu” Units of Excellence 
program MDM-2016-0692 and the Spanish Research State Agency.
KO, TR and AW were supported by the Polish National Science Centre, 
grant NCN 2019/35/B/ST2/00059.

We also thank Nick Manton for reading the manuscript and for calling our attention to Ref. \cite{kev}.


\begin{thebibliography}{99}
\bibitem{footnote1}
{It turns out that for this model the KAK separation parameter  $s$ is not a globally valid moduli space coordinate. Concretely, there exists a simple relation between $\phi_0$ and $s$ for $\phi_0 \ge -1$, but there is no sensible definition for $s$ for $\phi_0 <-1$.}
\end{thebibliography}
\end{document}